\documentstyle[epsfig]{elsart}
\newbox\mybox
\newcommand\fverb{\setbox\mybox=\hbox\bgroup\verb}
\newcommand\fverbdo{\egroup\medskip\noindent\fbox{\unhbox\mybox}\ }
\newcommand\fverbit{\egroup\item[\fbox{\unhbox\mybox}]}

\newcommand\init[1]{\setbox\mybox=\hbox{{\beeg #1}~}%
                   \noindent\global\hangindent=\wd\mybox\global\hangafter-2%
                   \sc\smash{\llap {\lower 13.2pt \box\mybox}}}
\begin{document}
\begin{frontmatter}

%
\title{Mott effect at the chiral phase transition and anomalous $J/\psi$ 
suppression}
\author[ur]{Gerhard R.G. Burau,}
\author[ur]{David B. Blaschke,}
\author[dub]{and Yuri L. Kalinovsky}
\address[ur]{Fachbereich Physik, Universit\"at Rostock, D-18051 Rostock,
Germany}
\address[dub]{Laboratory of Information Technologies, JINR, 141980 Dubna, 
Russia}
\begin{abstract}
We investigate the in-medium modification of the charmonium break-up 
process due to the Mott effect for light ($\pi$) and open-charm ($D$, $D^*$) 
mesons  at the chiral/deconfinement phase transition.
A model calculation for the process $J/\psi+\pi\to D+\bar D^* + h.c.$ 
is presented which demonstrates that the Mott effect for the D-mesons leads to 
a threshold effect in the thermal averaged break-up cross section. 
This effect is suggested as an explanation of the phenomenon of anomalous 
$J/\psi$ suppression in the CERN NA50 experiment.
\end{abstract}
\begin{keyword}
$J/\psi$ suppression, bound state dissociation, Mott effect\\[2mm]
%
\end{keyword}
\end{frontmatter}
%
\section{Introduction} 
Recent results of the CERN NA50 collaboration on anomalous $J/\psi$ 
suppression \cite{na50} in ultrarelativistic Pb-Pb collisions at 158 AGeV 
have renewed the quest for an explanation of the processes which may cause the 
rather sudden drop of the $J/\psi$ production cross section for transverse 
energies above $E_T\sim 40$ GeV in this experiment.
An effect like this was predicted as a signal for quark gluon plasma 
formation \cite{ms} due to screening of the $c \bar{c}$ interaction.
Soon after that it became clear that for temperatures and densities just above
the deconfinement transition the Mott effect for the $J/\psi$ does not occur
and that a kinetic process is required to dissolve the $J/\psi$ \cite{b} in a 
break-up process by impact of thermal photons \cite{Hansson:1988uk}, 
quarks \cite{rbs}, gluons \cite{Kharzeev:1994pz} 
or mesons \cite{Vogt:1988fj,mbq}.

In this paper, we suggest that at the chiral/deconfinement phase transition 
the charmonium break-up reaction cross sections are critically enhanced 
since the light and open-charm mesonic states of the dissociation processes 
become unbound (Mott effect) so that the reaction thresholds are effectively 
lowered.
We present a model calculation for the particular process 
$J/\psi+\pi\to D+\bar D^* + h.c.$ in a hot gas of resonant (unbound but 
correlated) quark-antiquark states in order to demonstrate that the Mott 
dissociation of the final states (D-mesons) at the chiral phase transition
leads to a threshold effect for the in-medium $J/\psi$ break-up cross section 
and thus the survival probability.

\section{In-medium modification of charmonium break-up cross sections}

The inverse lifetime of a charmonium state in a hot and dense many-particle 
system is given by the imaginary part of its selfenergy 
\begin{equation}
\tau^{-1}(p) = \Gamma(p) = \Sigma^>(p) - \Sigma^<(p)~.
\end{equation}
%
\begin{figure}[htb]
\centering{\
\epsfig{figure=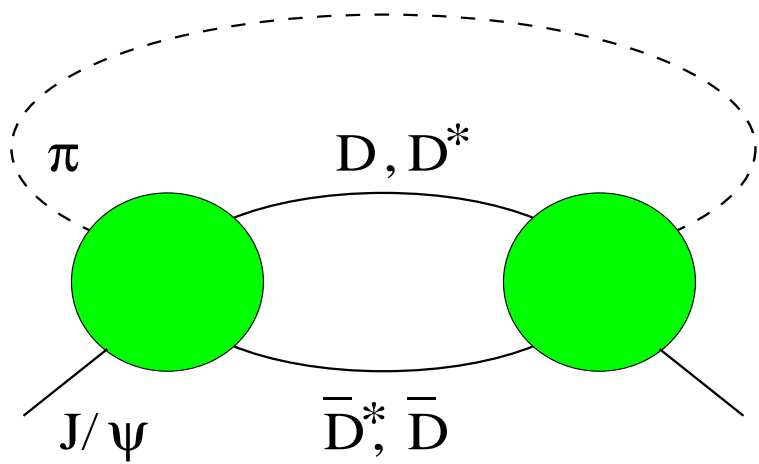,height=6.0cm,width=10.0cm}}
\caption{Diagrammatic representation of the complex selfenergy
for the $J/\psi$ due to break-up in (off-shell) $D$, $\bar{D}^*$ pairs by 
impact of (off-shell) pions from a hot medium.
\label{fig:cex}}
\end{figure}

In the Born collision approximation for the dominant process in a hot gas of 
pion-like correlations, as shown in Fig. \ref{fig:cex}, we have \cite{kb}
\begin{equation}
\Sigma^{\stackrel{>}{<}}(p) = \int\limits_{p'} 
\int\limits_{p_1}\int\limits_{p_2} (2\pi)^4 \delta_{p,p';p_1,p_2} 
\left|{\cal M}\right|^2 
G^{\stackrel{<}{>}}_{\pi}(p')~ 
G^{\stackrel{>}{<}}_{D_1}(p_1)~ 
G^{\stackrel{>}{<}}_{D_2}(p_2)~,
\end{equation}
where the thermal Green functions 
$G^{>}_{i}(p) = [1 + f_i(p)] A_i(p)$ and $G^{<}_{i}(p) = f_i(p) A_i(p)$ 
are defined by the spectral function $A_i(p)$ and the distribution function
$f_i(p)$ of the bosonic state $i$; with the notation 
$\int_p = \int \frac{d^4p}{(2\pi)^4}$, 
$\delta_{p,p';p_1,p_2} = \delta(p + p' - p_1 - p_2)$. 

In the low density approximation for the final states ($f_{D_i}(p) \approx 0$),
one can safely neglect $\Sigma^<(p)$ so that 
\begin{equation}
\tau^{-1}(p) = \int\limits_{p'} \int\limits_{p_1} \int\limits_{p_2} 
(2\pi)^4 \delta_{p,p';p_1,p_2} 
\left| {\cal M} \right|^2 
f_\pi(p')~ A_\pi(p')~ A_{D_1}(p_1)~ A_{D_2}(p_2). 
\end{equation}
With the differential cross section 
\begin{equation}
\frac{d\sigma}{dt} = \frac{1}{16\pi} 
\frac{\left| {\cal M}(s,t) \right|^2}
{\lambda(s,M_\psi^2,s')}~, 
\end{equation}
using $s = (p + p')^2$, $t = (p - p_1)^2$, $s' = p'^2$ and 
$\lambda(s, M_{\psi}^2, s') = 
[s - (M_{\psi} + \sqrt{s'})^2][s - (M_{\psi} - \sqrt{s'})^2] 
= 4~ v^2_{\rm rel}~ [{\bf p}^2 + M_{\psi}^2][{\bf p'}^2 + s']$
one can show that the $J/\psi$ relaxation time in a hot pion as well as pionic 
resonance gas is given by 
\begin{equation}
\label{tau}
\tau^{-1}(p) = \int \frac{d^3{\bf p'}}{(2\pi)^3} 
\int ds' f_\pi({\bf p'},s')~ A_\pi(s') 
v_{\rm rel}~ \sigma^*(s)~, 
\end{equation}
where depending on the properties of the medium the pion spectral function 
describes either $q \bar{q}$ bound states or resonant (off-shell) 
correlations. The in-medium break-up cross section is given by
\begin{equation}
\label{sig*}
\sigma^*(s) = \int ds_1~ ds_2~ 
A_{D_1}(s_1)~ A_{D_2}(s_2)~ \sigma(s; s_1, s_2)~. 
\end{equation}
Note that there are two kinds of medium effects due to (i) the spectral 
functions of the final states and (ii) the explicit medium dependence of the 
matrix element ${\cal M}$. 
In the following model calculation we will use the approximation 
$\sigma(s; s_1, s_2) \approx \sigma^{\rm vac}(s; s_1, s_2)$ 
justified by the locality of the transition matrix ${\cal M}$ which makes it 
rather inert against medium influence.

\section{Model calculation}

The quark exchange processes in meson-meson scattering can be calculated 
within the diagrammatic approach of Barnes and Swanson \cite{bs} which allows 
a generalization to finite temperatures in the thermodynamic Green function
technique \cite{br}. 
This technique has been applied to the calculation of $J/\psi$ break-up cross
sections by pion impact in \cite{mbq}. 
The approach has been extended to excited charmonia states and consideration 
of rho-meson impact recently \cite{wsb}.
The generic form of the resulting cross section (given a band of uncertainty)
can be fit to the form 
\begin{equation}
\label{sig0}
\sigma^{\rm vac}(s; M_{D_1}^2, M_{D_2}^2) = 
\sigma_0 \ln(s/s_0) \exp(-s/\lambda^2)\quad ,\quad s \ge s_0~,
\end{equation}
where $s_0 = (M_{D_1} + M_{D_2})^2$ is the threshold for the process to occur,
$\sigma_0 = 7.5 \cdot 10^{9}$ mb and $\lambda=0.9$ GeV.

Recently, the charmonium dissociation processes have been calculated also in
an effective Lagrangian approach \cite{mm,haglin}, but the freedom of choice 
for the formfactors of meson-meson vertices makes predictions uncertain. 
The development of a unifying approach on the basis of a relativistic confining
quark model is in progress \cite{kb99} and will remove this uncertainty by 
providing a derivation of the appropriate formfactors from the underlying 
quark substructure. 

The major modification of the charmonium break-up process which we expect at 
finite temperatures in a hot medium of strongly correlated quark-antiquark 
states comes from the Mott effect for the light as well as the open-charm 
mesons. 
At finite temperatures when the chiral symmetry in the light quark sector is 
restored, the continuum threshold for light-heavy quark pairs drops below the 
mass of the D-mesons so that they are no longer bound states constrained
to their mass shell, but become rather broad resonant correlations in the 
continuum. 
This Mott effect has been discussed within relativistic quark models 
\cite{b+93} for the light meson sector but can also be generalized to the case 
of heavy mesons 
\cite{gk}. Applying a confining quark model \cite{Blaschke:1998gk} we have 
obtained the critical temperatures $T^{\rm Mott}_{D^*}=110$ MeV, 
$T^{\rm Mott}_{D}=140$ MeV and $T^{\rm Mott}_{\pi}=150$ MeV 
\cite{Blaschke:2000er}. 

\begin{figure}[htb]
\centering{\
\epsfig{figure=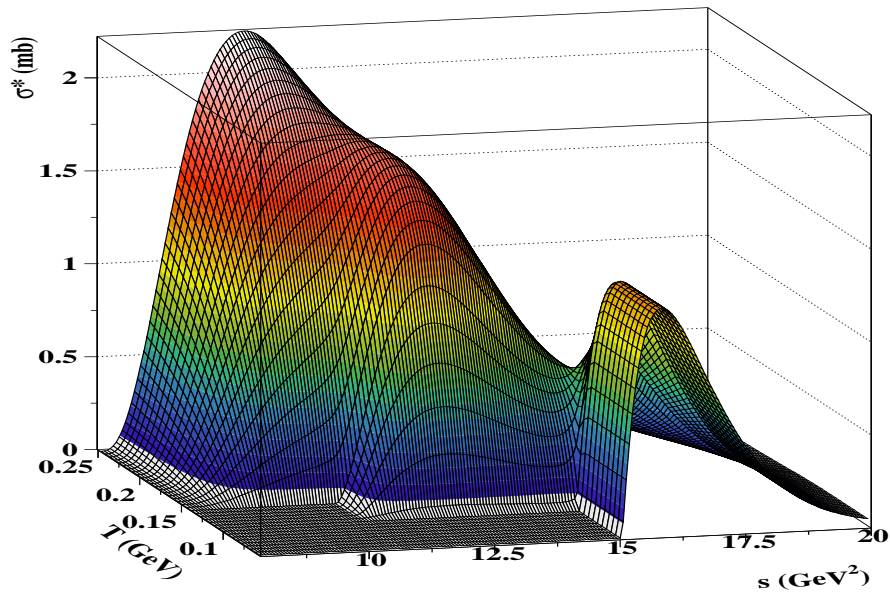,height=10.0cm,width=12.0cm}}
\vspace{-1cm}
\caption{Energy- and temperature dependent in-medium $J/\psi$ break-up cross 
section for pion impact. Thresholds occur at the Mott temperatures for the 
open-charm mesons: $T^{\rm Mott}_{D^*} = 110$ MeV, $T^{\rm Mott}_{D} = 140$ 
MeV.
\label{fig:sig_t}}
\end{figure}

In order to study the implications of the pion and $D$-meson Mott effect for 
the charmonium break-up we adopt here a Breit-Wigner form for the spectral 
functions 
\begin{eqnarray}
\label{ad}
A_i(s) &=& \frac{1}{\pi} 
\frac{\Gamma_i(T)~M_i(T)}{(s - M_i^2(T))^2 + \Gamma_i^2(T) M_i^2(T)}~,
\end{eqnarray} 
which in the limit of vanishing width $\Gamma_i(T)\to 0$ goes over into the 
delta function $\delta(s - M_i^2)$ for a bound state in the channel $i$.
The width of the pions as well as the $D$-mesons shall be modeled by a 
microscopic approach. For our exploratory calculation, we adopt here
\begin{equation}
\Gamma_{\pi,D}(T) = c~ (T - T^{\rm Mott}_{\pi,D})~ \Theta(T - T^{\rm Mott}_{\pi,D})~,
\end{equation}
where the coefficient $c = 2.67$ is assumed to be universal for the pions and 
$D$-mesons and it is obtained from a fit to the pion width above the pion Mott 
temperature, see \cite{b+95}. For the meson masses we have 
$M_{\pi,D}(T) = M_{\pi,D} + 0.75~ \Gamma_{\pi,D}(T)$.
The result for the in-medium $J/\psi$ break-up cross section (\ref{sig*}) is 
shown in Fig. \ref{fig:sig_t}.

With $M_{D^*}=2.01$ GeV and $M_{\bar D}=1.87$ GeV follows for the threshold
$s_0 = 15.05$ GeV$^2$. At a temperature $T = 140$ MeV, where the D-meson can 
still be considered as a true bound state, the $D^*$-meson has already entered 
the continuum and is a resonance with a half width of about 80 MeV.
Due to the Mott effect for the open-charm mesons (final states), 
the charmonium dissociation processes become "subthreshold" ones and their 
cross sections which are peaked at threshold rise and spread to lower onset 
with cms energy. 
This is expected to enhance strongly the rate for the charmonium dissociation 
processes in a hot resonance gas.

\section{$J/\psi$ dissociation in a hot ``pion'' gas}

We calculate the inverse relaxation time for a $J/\psi$ at rest 
in a hot gas of pions (below $T_{\pi}^{\rm Mott}$) and pion-like $q \bar{q}$ 
correlations (above $T_{\pi}^{\rm Mott}$) by specifying Eq. (\ref{tau}) for 
this case 
\begin{eqnarray}
\tau^{-1}(T) &=& \int\frac{d^3{\bf p'}}{(2\pi)^3}
\int ds_{\pi} A_{\pi}(s_{\pi}) 
f_\pi({\bf p'}, s_{\pi}; T) \frac{|{\bf p'}|}{E_\pi({\bf p'}, s_{\pi})} 
\sigma^*(s)\\ \nonumber \\
&=& <\sigma^* v_{\rm rel}>n_\pi(T)~, 
\end{eqnarray}
with the dispersion relation 
$E_\pi({\bf p'}, s_{\pi}) = \sqrt{{\bf p'}^2 + s_{\pi}}$, 
the thermal Bose distribution function 
$f_\pi({\bf p'}, s_{\pi}; T) = 3 \left\{\exp[E_{\pi}({\bf p'}, s_{\pi})/T] - 1\right\}^{-1}$ 
and the particle density $n_{\pi}(T)$ for the ``pions''. The cms energy of the 
``pion'' impact on a $J/\psi$ at rest is $s({\bf p'}; s_{\pi}) = s_{\pi} + M_{\psi}^2 + 2M_{\psi}E_\pi({\bf p'}, s_{\pi})$. 

\begin{figure}[htb]
\centering{\
\epsfig{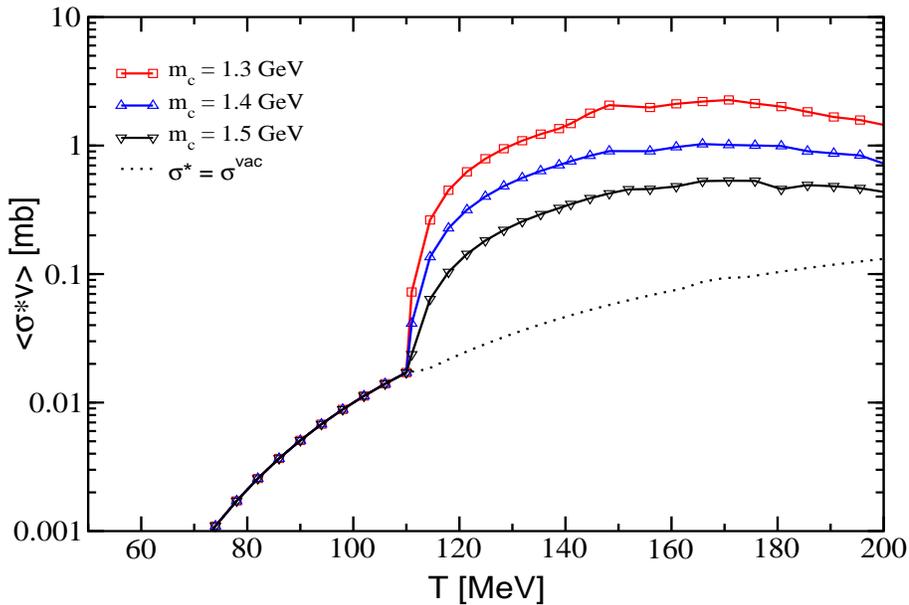}}
\caption{Temperature dependence of the thermal averaged in-medium $J/\psi$ 
break-up cross section for different charm quark masses.
\label{fig:sigv}}
\end{figure}

The result for the temperature dependence of the thermal averaged $J/\psi$ 
break-up cross section $<\sigma^* v_{\rm rel}>$ is shown in 
Fig. \ref{fig:sigv}. 
This quantity has to be compared to the nuclear absorption cross section for 
the $J/\psi$ of about 3 mb which has been extracted from charmonium 
suppression data in p-A collisions \cite{hhk}.
\vspace{0.5cm}

It is remarkable that it is practically negligible below the D-meson Mott
temperature $T^{\rm Mott}_{D^*}=110$ MeV but comparable to the nuclear 
absorption cross section above the chiral/deconfinement temperature of
$T_{\rm crit}\approx 150$ MeV. It is obvious that the transition from D-meson 
bound states to unbound light-heavy quark correlations is responsible for the 
strong increase by one to two orders of magnitude. 
Note that in this calculation the Mott effect for the pion (initial state) 
above $T^{\rm Mott}_{\pi}$ has been included, but does not alter the result 
obtained previously \cite{Blaschke:2000er} in a calculation neglecting this 
effect.

Therefore we expect the in-medium enhanced charmonium dissociation process  
to be sufficiently effective to destroy the charmonium state on its way 
through the hot fireball of the heavy-ion collision and to provide an 
explanation of the observed anomalous $J/\psi$ suppression phenomenon 
\cite{na50}. 
A detailed comparison with the recent data from the NA50 collaboration
requires a model for the heavy-ion collision. 
The effective in-medium break-up cross section for the $J/\psi$ derived in 
this work provides an input for all calculations which use this quantity, 
e.g. Glauber-type models 
\cite{Blaschke:2000er,wong,mb,Capella:2000zp,Blaizot:2000ev,Hufner:2000nv}, 
more detailed calculations based on a parton cascade model \cite{Bass:1999vz} 
or molecular dynamics \cite{Cassing:2000bj}.

\section{Summary and Outlook}

In this letter we have presented an approach to charmonium break-up 
in a hot and dense medium which is applicable in the vicinity of the 
chi\-ral/de\-con\-fine\-ment phase transition where mesonic bound states get 
dissolved in a Mott-type transition and should be described as resonant 
correlations in the quark plasma. This description can be achieved using the 
concept of the spectral function which can be obtained from relativistic quark 
models in a systematic way. The result of an exploratory calculation employing 
a temperature-dependent Breit-Wigner spectral function for light and 
open-charm mesons presented in this paper has demonstrated that heavy-flavor 
dissociation processes are critically enhanced at the QCD phase transition and 
could represent the physical mechanism behind the phenomenon of anomalous 
$J/\psi$ suppression.

In subsequent work we will relax systematically approximations which have been 
made in the present paper and improve inputs which have been used. 
In particular, we will investigate the off-shell behaviour of the charmonium 
break-up cross section in the vacuum (\ref{sig0}) and calculate the spectral 
functions (\ref{ad}) at finite temperature within a relativistic quark model.
Dyson-Schwinger equations provide a nonperturbative, field-\-theo\-retical 
approach which has recently been applied also to heavy-meson observables 
\cite{ikr} and have proven successful for finite-temperature generalization 
\cite{bbkr,Roberts:2000aa}. 
Further intermediate open-charm states can be considered; the states in the
dense environment should include rho mesons and nucleons besides of the 
pions which all can be treated as off-shell quark correlations at the QCD 
phase transition. 

In future experiments at LHC the charm distribution in the created fireball 
may be not negligible so that the approximation $f_{D_i}(p) \approx 0$ has to 
be relaxed. In this case, one has to include the gain process ($D \bar{D}$ 
annihilation) encoded in the $\Sigma^{<}$ function. 
In comparision to previous investigations \cite{ko,pbm} the present quantum 
kinetic treatment contains Bose enhancement factors in the $G^{>}_{i}$ 
functions which modify the charm equilibration process.

\section{Acknowledgements}

This work has been supported by the Heisenberg-Landau program for scientific 
collaborations between Germany and the JINR Dubna and by the DFG 
Graduiertenkolleg ``Stark korrelierte Vielteilchensysteme'' at the University 
of Rostock. We thank T. Barnes, P. Braun-Munzinger, J. H\"ufner, M.A. Ivanov, 
C.D. Roberts, G. R\"opke, S.M. Schmidt and P.C. Tandy for their discussions
and stimulating interest in our work.


\end{document}